# Type-II Superconductivity below 4K in $Sn_{0.4}Sb_{0.6}$

M.M. Sharma[1,2], Kapil Kumar[1,2], Lina Sang[3], X.L. Wang[3] and V.P.S. Awana[1,2, *]

[1]*Academy of Scientific & Innovative Research (AcSIR), Ghaziabad-201002*
[2]*CSIR- National Physical Laboratory, New Delhi-110012*
[3]*Institute of superconducting and electronic materials, University of Wollongong, NSW 2522, Australia*

**Abstract**

In this article, we report the occurrence of superconductivity in $Sn_{0.4}Sb_{0.6}$ single crystal at below 4K. Rietveld refined Powder XRD data confirms the phase purity of as grown crystal, crystallizing in rhombohedral R-3m space group with an elongated (2xc) unit cell in c-direction. Scanning Electron Microscope (SEM) image and EDAX measurement confirm the laminar growth and near to desired stoichiometry ratio. Raman Spectroscopy data shows the vibrational modes of Sn-Sb and Sb-Sb modes at 110 and 135cm$^{-1}$. ZFC (Zero-Field-Cooled) magnetization measurements done at 10Oe showed sharp superconducting transitions at 4K along with a minor step at 3.5K. On the other hand, Paramagnetic Meissner Effect (PME) is observed in FC measurements. Magnetization vs applied field (M-H) plots at 2, 2.2, 2.5, 2.7, 3, 3.2, 3.5, and 3.7K shows typical Type-II nature of observed superconductivity with lower and upper critical fields ($H_{c1}$ and $H_{c2}$) at 69.42Oe and 630Oe respectively at 2K. Type-II superconductivity is also confirmed by calculated Ginzburg-Landau Kappa parameter value of 3.55. Characteristics length viz. coherence length and penetration depth are also calculated. Weak granular coupling is observed from ρ-T plot, in which resistance is not dropping to zero down to 2K.

*Corresponding Author

Dr. V. P. S. Awana: E-mail: awana@nplindia.org
Ph. +91-11-45609357, Fax-+91-11-45609310
Homepage: awanavps.webs.com

**Introduction**

Presently, the condensed matter scientists are mostly busy studying new quantum phases of the matter, including the topological insulators (TI). TIs are highly appreciated for their unique properties due to existence of gapless surface states along with the insulating bulk [1-3]. Topological insulators, when accompanied with superconductivity, form an interesting class of materials called Topological Superconductors (TS). TS does show some distinct features such as hosting Majorana fermions [4], which are used in fault tolerant quantum computers [5]. Doping is a well-established method to induce superconductivity in bulk TIs like $Bi_2Se_3$, $Bi_2Te_3$ as $Bi_2Se_3$, which care made superconducting by intercalation of Copper (Cu), Niobium (Nb), Strontium (Sr), Thallium (Tl) and Palladium (Pd) [6-10]. There are some binary alloys as well; containing heavy elements such as Bismuth (Bi) that shows topological non trivial behavior due to the presence of intrinsic spin orbit coupling (SOC) in these elements [11-12]. $Bi_{1-x}Sb_x$ falls in this category of TIs and is also the first observed 3D TI [13]. So it is important to study the superconductivity in these topologically non trivial binary alloys. In this context, binary alloy containing Sn and Sb draws the attention as Sb also have topological non trivial surface states [14,15], hence its alloy may show topological behavior. This system of Sn and Sb is structurally different from that of $Bi_{1-x}Sb_x$ in which there is a random distribution of atoms of Bi and Sb in lattice. In case of $Sn_{1-x}Sb_x$ lattice, the structure is made up of alternating layers of Sn and Sb atoms [16]. Superconductivity in SnSb is reported a long time ago [17], in a joint report of intermetallic superconductors. The interest renewed once again when recently, it is been predicted from first principal density functional calculations [15] that same may be a TS and a detailed report showing bulk superconductivity in SnSb appeared in 2018 [18]. In a recent study it has been shown that critical temperature can be tuned in SnSb by changing the elemental composition viz $Sn_{1-x}Sb_x$.

In regards of topological nature of $Sn_{1-x}Sb_x$, first principles calculation in Ref. 15 shows considerable band opening on inclusion of SOC. It is a known fact that topological properties in materials can be triggered by hetero-structure engineering [19], which involves modification in stacking sequence of atomic layers or insertion of extra atomic layers in unit cell. The alteration of topological properties by this hetero structure engineering is evident from large difference in topological properties of $(Sb_2)_m$-$(Sb_2Te_3)_n$ where $Sb_2$ bilayer are inserted between quintuple layers of $Sb_2Te_3$ [20]. Here the constituent elements of $Sn_{1-x}Sb_x$,

Sn and Sb atoms both are supposed to have non-trivial surface states [14, 21] and these are stacked in unit cell as alternate atomic layers thus making it an example of natural super lattice structures. This creates a possibility for $Sn_{1-x}Sb_x$ to show topological properties and with its superconducting properties it can be considered as a possible candidate of topological superconductors. Because, this claim does not have sufficient experimental evidence, hence it is important to study more about its superconducting and topological properties.

In this paper we report the structural and superconducting properties of $Sn_{0.4}Sb_{0.6}$ which may be a candidate of topological superconductivity. From Rietveld refinement of Powder XRD pattern, we found clear evidence that two phases of $Sn_{1-x}Sb_x$ are present with two different unit cells; one with single and other with elongated (2xc) along c-axis. This approach of two unit cells is not yet reported in literature for this system. DC magnetization measurements exhibited superconductivity below 4K. In an earlier report [15], superconductivity is observed at below 4K for x=0.43, albeit with a relatively broader transition. M-H plots at various temperatures give a clear glimpse of Type-II superconductivity in this material, neglecting the possibility that this enhanced $T_c$ may be due to Sn, which is Type-I superconductor with $T_c$ near to 3.7K.

**Experimental**

$Sn_{0.4}Sb_{0.6}$ crystal is synthesized by solid state reaction route via vacuum encapsulation in PID controlled muffle furnace. 4N pure Sn and Sb powder were taken in stoichiometry ratio and ground thoroughly by agate mortar pestle for 30 minutes in MBRAUN glove box filled with argon gas. The grounded powder was then palletized with the help of hydraulic pressure palletizer at a pressure of 50 gm/cm$^3$. After that this palletized sample was vacuum sealed in a quartz ampoule at a pressure of $5*10^{-5}$ Torr and then placed into PID controlled muffle furnace. The vacuum sealed sample was melted by heating to 900$^0$C at a rate of 120$^0$C/hour and this melted sample was kept at 900$^0$C for 48 hours to make the melt homogenous. Then melted sample was cooled slowly to 300$^0$C at a rate of 2$^0$C/h so that the atoms attain their lowest energy positions to meet crystallinity. Phase diagram of $Sn_{1-x}Sb_x$ shows that $Sn_{0.4}Sb_{0.6}$ phase stabilizes at 300$^0$C [22] hence the sample was annealed at 300$^0$C for 72 hours and then directly quenched into ice water to avoid formation of any low temperature phase with different stoichiometry. The grown crystal looked silvery shiny and easily cleavable. The schematic of heat treatment, representing all steps of growth viz. heating, hold at high temperature, slow cooling, annealing and quenching of as grown $Sn_{0.4}Sb_{0.6}$ is shown in Fig.1 and image of flat surface of grown crystal being mechanically cleaved along its growth axis is shown in inset of the same Fig. Here Rigaku made Mini Flex II X-ray diffractometer having $CuK_\alpha$ radiation of 1.5418Å wavelength of X-rays is used to obtain PXRD data of powder of as grown $Sn_{0.4}Sb_{0.6}$ single crystal. Rietveld refinement of powder XRD data is done by using Full Proof Software. VESTA software is used to draw unit cell based on refined parameters from Rietveld analysis to visualize atomic structure of the as grown $Sn_{0.4}Sb_{0.6}$ single crystal. Field Emission Scanning Electron Microscope (FESEM) is used to visualize the morphology of as grown $Sn_{0.4}Sb_{0.6}$ single crystal. Presence of all constituent elements in desired stoichiometry ratio of as grown $Sn_{0.4}Sb_{0.6}$ single crystal is confirmed by Energy Dispersive X-ray Analysis (EDAX). Raman spectra for as grown $Sn_{0.4}Sb_{0.6}$ single crystal are recorded by using Renishaw Raman Spectrometer. All electrical and magnetic measurements of $Sn_{0.4}Sb_{0.6}$ single crystal were performed on QD-PPMS (Physical Property Measurement System). Four probe conventional method is used for Resistivity vs Temperature measurements on PPMS down to 2K.

**Results & Discussion**

Rietveld refinement of PXRD data of as grown $Sn_{0.4}Sb_{0.6}$ single crystal is shown in Fig. 2(a), which shows that the as grown crystal is crystallized in rhombohedral structure with R-3 m space group. Both Sn and Sb occupy same atomic positions viz. (0, 0, 0). Refined lattice parameters are a = b = 4.3267(8)Å & c = 5.3409(1)Å, having $\alpha = \beta = 90^0$ and $\gamma = 120^0$. Quality of fit can be determined by various parameters such as R- values and $\chi^2$ (goodness of fit parameter). Most relevant is R-factor i.e., $R_{exp}$, which is used to determine $\chi^2$. In our case $R_{exp}$ is 7.46 and parameter of goodness of fit $\chi^2$ is found to be 5.45. These fit parameters are in acceptable range, despite of being having tiny unfitted superstructures peaks. In PXRD pattern given in ref. 18, peak splitting is observed at $51.5^0$, and is attributed as signature of distorted rhombohedral structure. Here we find no splitting in XRD peak observed at this angle as shown in right hand side inset of Fig. 2(a). On the other hand, there are two un- split peaks at $51.5^0$ and both are fitted and indexed as (003) and (021) plane reflections without any superstructure. Left hand side inset of Fig. 2(a) shows the presence of extra unfitted peaks at $40^0$ and $42.5^0$, which indicate the sign of presence of superstructures or another lattice within as grown crystal. These peaks could not be fitted by single unit cell approach, even after introducing the superstructure [18]. Hence, the concept of two unit cells is used to fit the observed peaks. It is evident in literature that doping of impurities in sample can lead to occurrence of two unit cells having same space group; one with single unit cell and other with elongated unit cell along c-axis. This situation is more abundant when the host and impurity level is close to 50% each. This is observed in case of Al doping in $MgB_2$ where c parameter of $MgB_2$ unit cell is doubled in c-direction by doping of Al impurity at 40% level [23]. To confirm the presence of double unit cell Rietveld refinement is performed for two phase.

Fig. 2 (b) shows Rietveld refinement of Powder XRD data of as grown $Sn_{0.4}Sb_{0.6}$ single crystal in two phases, one with c-parameter of single cell and other with elongated unit cell along c axis. The elongated c-parameter is almost doubled to that as obtained for single unit cell refinement. The values of lattice parameters obtained from both refinements are a=b= 4.2762Å & c=11.4097Å, $\alpha = \beta = 90^0$ and $\gamma=120^0$; the ration c/a is increased from 1.234 to 2.668. Fitting parameters i.e., $R_{exp.}$ and parameter of goodness of fit $\chi^2$ are found to be 7.93 and 4.54 respectively without background calculation, all these parameters are in acceptable range and shows quality of fit to be reasonably good. This refinement shows the presence of reflections for both phases having single and elongated (2xc) unit cells simultaneously and their percentage is 66.7% and 33.3% respectively. It is shown in inset of Fig. 2(c) that the super structural peaks observed at $40^0$ and $42.5^0$ are well fitted by using the concept of elongated unit cell and these peaks are indexed as (104) and (110) plane reflections. The reason behind the elongation of unit cell along c axis is the structure of $Sn_{1-x}Sb_x$, which are lamellae of alternating atomic layers of Sn and Sb atoms (Sn-Sb-Sn-Sb-Sn-Sb-Sn). Lattice parameters along with goodness of fitting for both single/double unit cell concepts are given in Table 1. Fig. 2(c) depicts the unit cell structure of as grown $Sn_{0.4}Sb_{0.6}$ drawn by using VESTA software based on parameters obtained from Rietveld refinement with elongated unit cell. This shows the alternating atomic layers of Sn and Sb atoms. From this structure, it is clear that when Sb atoms are introduced into face centered cubic unit cell of Sn, the same does not simply replace the Sn atoms as in $Bi_{1-x}Sb_x$ where Sb atoms replaces Bi atoms and lead to random distribution of atoms. Here in case of $Sn_{1-x}Sb_x$, Sb atoms fills the voids that are present in Sn unit cell and creates their own atomic positions in unit cell as shown in Fig. 2(c). When the doping level is gradually increased, its structure and cell parameters also begin to change. As the doping level reaches beyond 50%, the structure becomes completely rhombohedral and unit cell becomes elongated along c –axis due to insertion of several Sb atomic layers in unit cells. Here is it worth mentioning that detailed high resolution transmission electron microscopy (HRTEM) studies are warranted on the studied $Sn_{0.4}Sb_{0.6}$, to authentically ascertain the double unit cell concept as proposed by us from Rietveld refinement of PXRD.

Fig. 3(a) shows the Scanning Electron Microscope (SEM) image of studied $Sn_{0.4}Sb_{0.6}$ crystal with resolution of 2μm. This is taken to visualize the morphology of grown crystal. Usually SEM images of crystals, show stairs and terrace type morphology and as seen in Fig. 3, similar stairs and terraces are found for as grown $Sn_{0.4}Sb_{0.6}$ crystal. It is clear that studied $Sn_{0.4}Sb_{0.6}$ crystal is grown in layered form. EDAX spectrum of as grown $Sn_{0.4}Sb_{0.6}$ single crystal is shown in Fig. 3(b), with its elemental composition. Peaks observed for Sn and Sb in EDAX spectrum confirms the presence of all constituent elements viz. Sn and Sb. The two peaks observed in initial part of the spectra can be attributed to adventitious Carbon and Oxygen that are normally deposited on surface due to exposure of sample to atmosphere. Elemental composition obtained from EDAX is shown in attached table, which is found near to desired stoichiometry ratio.

To study the phonon dynamics and vibrational modes, Raman Spectroscopy on as grown $Sn_{0.4}Sb_{0.6}$ crystal is performed by using Ravenshaw Raman Spectrometer. The sample is irradiated with a Laser of 514nm, and 5mW power for 30 sec. The observed Raman spectrum is shown in Fig. 4. The spectrum is de-convoluted by using Lorentz fitting formula, which clearly shows the presence of three Raman Shift peaks at 70,110 and 135$cm^{-1}$. SnSb shows two prominent Raman shift peaks between at 110 and 150$cm^{-1}$ [24,25]. Hence forth, peaks at 110 and 135$cm^{-1}$ are attributed to the Raman vibrational modes of SnSb. Sb also shows two Raman active modes between 110 and 150$cm^{-1}$; one degenerate $E_g$ mode at 110$cm^{-1}$, which occurs due to the shifting of atoms in a direction perpendicular to $C_3$ axis and other $A_g$ mode at 150$cm^{-1}$, which occurs due to displacements of atoms along $C_3$ axis [26,27]. Here we find Raman active mode at 110$cm^{-1}$, which can be due to vibrations of both Sn-Sb and Sb-Sb, $E_g$ modes. Molecular masses of Sn and Sb are nearly same so there is not much difference between characteristic Raman shift of vibrations of Sn-Sb and Sb-Sb bonds. Surface driven Raman Shift peak observed near 70$cm^{-1}$ is of $Sb_2O_4$ [28], which might be formed due to the surface oxidation of as grown crystal during Raman Spectroscopy experiment as Laser irradiation could lead to formation of $Sb_2O_4$.

DC magnetization study below 6K under Zero Field Cooled (ZFC) and Field Cooled (FC) in presence of magnetic field of 10Oe is shown in Fig. 5. This M-T plot confirms the bulk superconductivity with $T_c^{onset}$ at 4K in as grown $Sn_{0.4}Sb_{0.6}$ crystal as a clear diamagnetic transition is appeared below this temperature in ZFC measurements. It is a noticeable result as in earlier report on this material (ref. 18), the bulk superconductivity is observed near 2K for x=0.6 composition, and only a minute transition was observed at 4K for x=0.43 composition, which was attributed to superconducting transition of Sn, but that is not the case here, the possibility that this transition can be due to un-reacted Sn is neglected from M-H plots, which will be discussed in later part. So observed $T_c$ in presently as grown $Sn_{0.4}Sb_{0.6}$ crystal is slightly higher to that is observed in earlier reports. Two superconducting transitions are visible in ZFC measurements; one at 4K and other tiny step at 3.5K, marked by arrow in M-T plot. The observed second transition may be due to presence of superconducting Sn, which shows Type-I superconductivity at 3.7K.

FC measurements shows paramagnetic Meissner effect below $T_c$ which is a different behavior to that as usually a superconductor shows [29], as diamagnetism below $T_c$ is supposed to be the hallmark of superconductivity. There are several mechanisms being purposed to explain this mysterious effect such as surface superconductivity [30, 31], vortex-vortex interaction [32, 33] and formation of pie junctions [34, 35]. Here the origin of observed Paramagnetic Meissner Effect (PME) can be attributed to the presence of two superconducting phases in sample because this effect arises due to surface superconductivity or due to presence of multiple superconducting phases that result in trapping of magnetic flux. Here paramagnetic Meissner effect is consistent with results of ZFC measurements as it show clear bifurcation of superconducting transition onset at below 4K. The 3.5K transition is not visible, because that is a very faint transition, even in ZFC magnetization. This faint transition is due to unreacted Sn, which is in normal state at 4K and act as a paramagnetic impurity. It is well studied that presence of paramagnetic impurity between two superconducting regions creates a phase difference of π angle in order parameters of two superconducting regions [36,37], this type of junctions are known as π junctions. From this π junction a spontaneous current is originated, which leads to trapping of magnetic flux near the junction. This inhomogeneous trapping of magnetic flux is responsible for the observed Paramagnetic Meissner effect. Similar kind of behaviour has been observed for $MgB_2$ where presence of unreacted Mg led to observed PME [29]. To get more insight about the observed PME time relaxation magnetic measurements are required to examine the behaviour of vortex and shall be the subject of separate article.

Magnetization vs applied field plots at different temperature viz. 2, 2.2, 2.5, 2.7, 3, 3.2, 3.5 and 3.7K are shown in Fig. 6. These wide open plots are consistent with the usual plots that are supposed to be observed for any Type-II superconductor and thus confirming the main superconducting phase to be Type-II. It can be clearly seen in these plots that upper and lower critical field are decreasing with increasing the temperature. In our sample superconducting transition is observed at higher temperature to that was reported earlier in ref. 18, so here question arises whether the observed superconducting transition is of Sn which shows Type-I superconductivity at 3.7K or it is the superconducting transition of $Sn_{0.4}Sb_{0.6}$. The open M-H loops right up to close to bulk superconducting transition at 4K indicates towards main phase $Sn_{0.4}Sb_{0.6}$. Further, the $Sn_{0.4}Sb_{0.6}$ M-H plots are clearly of Type-II superconductivity nature, while elemental Sn is a Type-I superconductor with a critical field of around 300Oe. It is clear from wide open M-H plots up to 3.7K (inset Fig.6) with $H_{c1}$ and $H_{c2}$ being 11Oe and 120Oe, that the observed superconductivity is not quite possible from elemental Sn. The M-H plots show clearly the presence of two critical fields viz. $H_{c1}$ and $H_{c2}$, this shows that the grown crystal is a Type-II superconductor.

Fig. 7(a) shows isothermal M-H plots for different temperatures in superconducting region. Linear response of magnetization with respect to low applied field is evident from Fig. 7(a) and it can be attributed as a clear signature of Meissner state. Lower critical field $H_{c1}$ is defined as the field at which M-H starts to deviate from linearity and enters to vortex state from Meissner state and M-H loop becomes highly irreversible. Several methods have been proposed till date to calculate $H_{c1}$ by different groups [38-43]. The one being purposed by Abdel-Hafiez et.al., in ref. 38, is among the most rigorous methods. In this method $H_{c1}$ is calculated by measuring the trapped magnetic flux $M_t$. Another, though not very accurate but popular method to determine $H_{c1}$ is by identifying the point where linear to nonlinear transition occurs in M-H plot. Here, the later method is applied to calculate $H_{c1}$. In this method, first the slope of low field magnetization curves is calculated by linear fit of magnetization curve at low field, this slop is used to calculate $M_0$. This obtained value of $M_0$ is then subtracted from each isotherm and plotted against applied magnetic field as shown in Fig. 7(b). $H_{c1}$ is determined by the point where ΔM vs H plot deviates from zero base line, this zero base line is shown by an arrow in Fig. 7(b). This is the field, where magnetic field starts to penetrate the sample. By above calculations the obtained values of $H_{c1}$ are 82, 73, 60, 49, 40, 33, 22 and 11Oe at 2K, 2.2K, 2.5K, 2.7K, 3.0K, 3.2K, 3.5K and 3.7K respectively. ΔM vs H plot at 2K is shown in inset of Fig. 7(b), where deviation of ΔM from zero base line is clearly visible at 82Oe. These experimentally obtained values of $H_{c1}$ need to be corrected to account demagnetization factors.

Demagnetization factor can be calculated by using the relation purposed by Brandt [44]. Demagnetization factor N can be calculated by relation

$$N = 1 - 1/(1+qa/b) \quad \ldots(1)$$

Where, a and b are dimension of the sample in perpendicular to field and along the field i.e. thickness of measured sample and these are 2.5mm and 0.20mm respectively. The value of q can be calculated by following formula

$$q = \frac{4}{3\pi} + \frac{2}{3\pi}\tanh\left[1.27\frac{b}{a}In\left(1+\frac{a}{b}\right)\right] \quad \ldots(2)$$

The obtained value of q is 0.46 and thus obtained value of demagnetization factor N from equation (1) is 0.8571. So corrected values of $H_{c1}$ are 69.42, 62.56, 50.826, 41.5079, 34.284, 28.28, 18.85 and 9.43Oe at 2K, 2.2K, 2.5K, 2.7K, 3.0K, 3.2K, 3.5K and 3.7K respectively. These corrected $H_{c1}$ values are plotted against temperature and shown in inset of Fig. 7(a), this shows that $H_{c1}$ decreases monotonically with increase in temperature. $H_{c1}$ is known to follow quadratic equation $H_{c1} = H_{c1}(0)*[1-T^2/T_c^2]$, the plot of $H_{c1}$ vs T is fitted with this equation (shown by black curve) and found to be in order as shown in inset of Fig. 7(a). The value of $H_{c1}(0)$ can be determined by extrapolating the fitted plot which is found to be 85Oe.

The corrected value of $H_{c1}$ at 2K is found to be 69.42Oe. Upper critical field $H_{c2}$ is marked in Fig. 7(a), where the M-H loop closes or touches the baseline, which coincides with the irreversibility of the field and its value is found to be 630Oe at 2K. Mean critical field $H_c$ of a Type-II superconductor can be calculated by using the formula Hc = $(H_{c1}*H_{c2})^{1/2}$, and it is found to be 209.12Oe. The value of upper critical field at 2K, $H_{c2}(T)$ is used to calculate upper critical field at absolute zero $H_{c2}(0)$ by using Ginzberg-Landau (G-L) equation which is given below:

$$H_{c2}(T) = H_{c2}(0) * \left[\frac{1-t^2}{1+t^2}\right]$$

In above equation t is reduced temperature and is written as t = $T/T_c$, T is the temperature at which experiment is carried out and $T_c$ is critical temperature. Here T is 2K and $T_c$ is 4K which gives the value of reduced temperature to be around 0.5. By putting the values of reduced tempearature t and upper critical field $H_{c2}(T)$ in given equation, $H_{c2}(0)$ is found to be 1050Oe. Type-II superconductivty in present sample can be confirmed from the value of G-L parameter or Kappa (κ) parameter as well. The value of κ parameter can be calculated by using the formula $H_{c2}(0) = \kappa*(2)^{1/2}*H_c$ and it is found to be 3.55, which is greater than threshold value for Type-I superconductivity i.e., $1/2^{1/2}$ thus confirming Type-II superconductivity in our as grown $Sn_{0.4}Sb_{0.6}$ crystal. Upper critical field at absolute zero $H_{c2}(0)$ is used to calculate the value of another superconducting critical coherence length $\xi(0)$ with the help of formula $H_{c2}(0) = \frac{\varphi_0}{2\Pi\xi(0)^2}$ , here $\phi_0$ is constant known as flux quanta and its value is $2.0678 \times 10^{-15}$ Wb. The calculated value of $\xi(0)$ from given equation is found to be 5.6Å. The value of coherence length $\xi(0)$ is used to estimate London penetration depth $\lambda(0)$ by using the realtion $\kappa = \lambda(0)/\xi(0)$. From this relation the value of $\lambda(0)$ is found to be 19.88Å.

Fig. 8 shows the resistivity vs temperature plot of as grown $Sn_{0.4}Sb_{0.6}$ crystal from 300K to 2K. The inset of this Fig. shows the extended $\rho$-T plot up to 5K. From the inset, it is clear that resistivity starts to decrease from 4K, this decrement in resistivity is marked by a straight line and a sharp transition is observed at 3K, which is in accordance to the two transitions observed in M-T plots. This plot gives the value of $T_c^{onset}$ as well above 3K. The $\rho$–T plot is fitted with equation $\rho = \rho_0 + A*T^2$ and is shown by solid black line, which is found to obey the same in normal state up to 55K, suggesting electron - electron type scattering in normal state. This is clear sign of Fermi liquid type behavior of as grown crystal in its normal state i.e., above superconducting transition. The value of $\rho_0$ is found to be around 83μohm-cm. Resistivity ratio $\rho_{300}/\rho_5$ found to be 1.44 which is low, suggesting that electron-electron type scattering is accompanied by disorder scattering due to antisite defect, which possibly arises due to alternate layers of Sn and Sb [18]. Resistance in $\rho$-T plot does not drop to zero down to 2K this can be due to weak granular coupling and presence of structural disorder as evident from low resistivity ratio of 1.44. This observed large superconducting transition width is due to presence of grains that are forming SIS or SNS junctions, which is in accordance to observed Paramagnetic Meissner effect in FC measurements. This weak coupling between grains leads to breaking of superconducting channels, which can be the reason that resistivity does not drop to zero down to lowest measured temperature i.e., 2K.

## Conclusion:

Summarily, we analyzed structural and superconducting properties of theoretically envisaged topological superconductor candidate $Sn_{0.4}Sb_{0.6}$. We proposed existence of elongated unit cell in $Sn_{0.4}Sb_{0.6}$ having a seven layer lamellae with alternating Sn and Sb layers forming septuple layers. Two bulk superconducting transitions; one major at below 4K and another very minor at 3.5K are observed. Paramagnetic Meissner effect is also observed due to presence of multiple superconducting crystalline layers in sample and their weak coupling is evident from ρ-T plot. Superconductivity critical parameters viz. G-L κ parameter, coherence length, penetration depth etc., are also calculated. It is clear that XRD (SnSb single/double unit phase), Raman (Sn-Sb modes) and magnetization viz. diamagnetic onset at 4K and wide Type-II superconductivity loop openings right up to 4K are all in support to the fact that the observed superconductivity is from $Sn_{0.4}Sb_{0.6}$ and not from elemental Sn.

**Acknowledgment**

Authors would like to thank Director CSIR-NPL for his keen interest and encouragement. Authors are also grateful to Mrs. Shweta for timely Raman Spectroscopy measurement. M.M. Sharma would like to thanks CSIR for research fellowship and AcSIR-Ghaziabad for PhD registration and Kapil Kumar would like to thank UGC for research fellowship and AcSIR for PhD registration.

**Table 1**

Lattice parameters table for $Sn_{0.4}Sb_{0.6}$ sample:

|  | a (Å) | b (Å) | c (Å) | Volume (Å$^3$) | c/a |
|---|---|---|---|---|---|
| Single cell Concept | 4.3267(8) | 4.3267(8) | 5.3409(1) | 99.987(1) | 1.234 |
| Double cell concept | 4.2762(9) | 4.2762(9) | 11.4097(6) | 208.646(3) | 2.668 |

**Figure Captions**

Fig.1: Schematic of heat treatment of sample of $Sn_{0.4}Sb_{0.6}$ and inset shows the image of a piece of grown crystal.

Fig.2: (a) Rietveld Refinement of Powder XRD pattern of $Sn_{0.4}Sb_{0.6}$.

(b) Rietveld Refinement of Powder XRD pattern of $Sn_{0.4}Sb_{0.6}$ in two phase one with single unit cell and other with elongated unit cell.

(c) Unit cell structure of grown $Sn_{0.4}Sb_{0.6}$ crystal drawn by using VESTA software.

Fig.3: (a) Field Emission Scanning Electron Microscope (FESEM) image of grown $Sn_{0.4}Sb_{0.6}$ crystal.

(b) EDAX spectra of grown $Sn_{0.4}Sb_{0.6}$ with elemental composition.

Fig.4: De-convoluted Raman Spectra of grown $Sn_{0.4}Sb_{0.6}$ taken at room temperature.

Fig.5: FC & ZFC measurements of grown $Sn_{0.4}Sb_{0.6}$ crystal in temperature range from 2K to 6K at a magnetic field of 10 Oe.

Fig.6: Isothermal magnetization (M-H) of $Sn_{0.4}Sb_{0.6}$ at various temperatures in superconducting regime i.e., between 2K to 3.7K, inset shows the extended plot at 3.7K.

Fig.7: (a) Expanded M-H plots of $Sn_{0.4}Sb_{0.6}$ at different temperature deep right up to superconductivity onset, marking clearly the $H_{c1}$ and $H_{c2}$, inset shows the fitted plot of $H_{c1}$ vs T.

(b) ΔM vs H plot of grown $Sn_{0.4}Sb_{0.6}$ crystal showing deviation of M from slop of low magnetic field data at temperature from 2K to 3.7K, inset shows the same at 2K marking the deviation from zero to determine $H_{c1}$ at 2K.

Fig.8: Resistivity versus temperature plot for studied $Sn_{0.4}Sb_{0.6}$, inset shows the expended plot of the same.

Fig.1

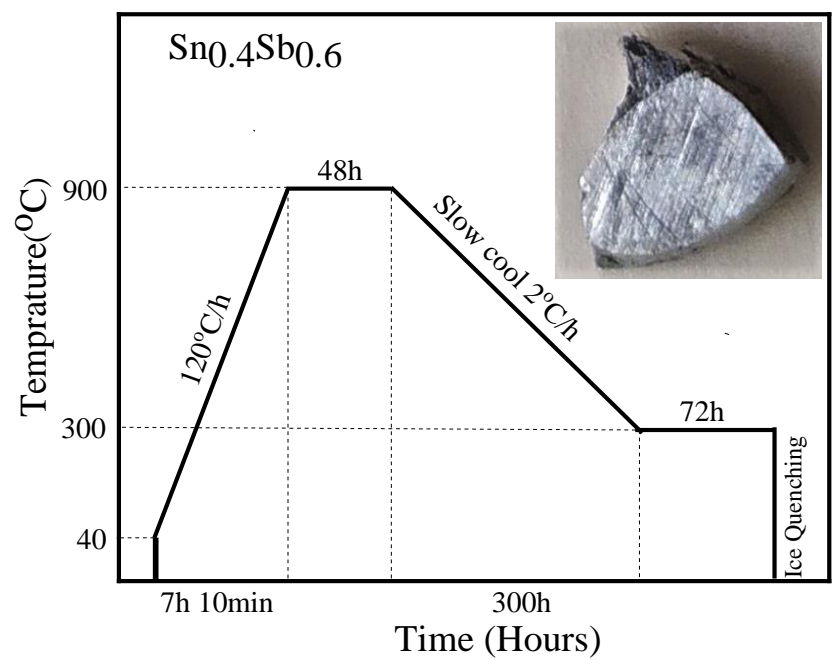

Fig. 2(a)

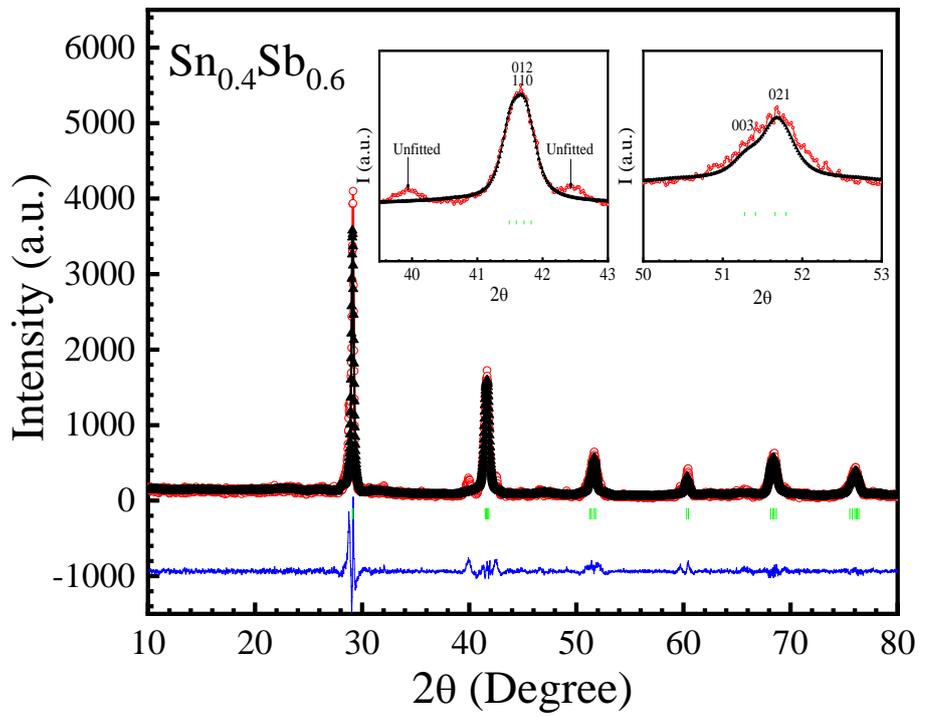

Fig. 2(b)

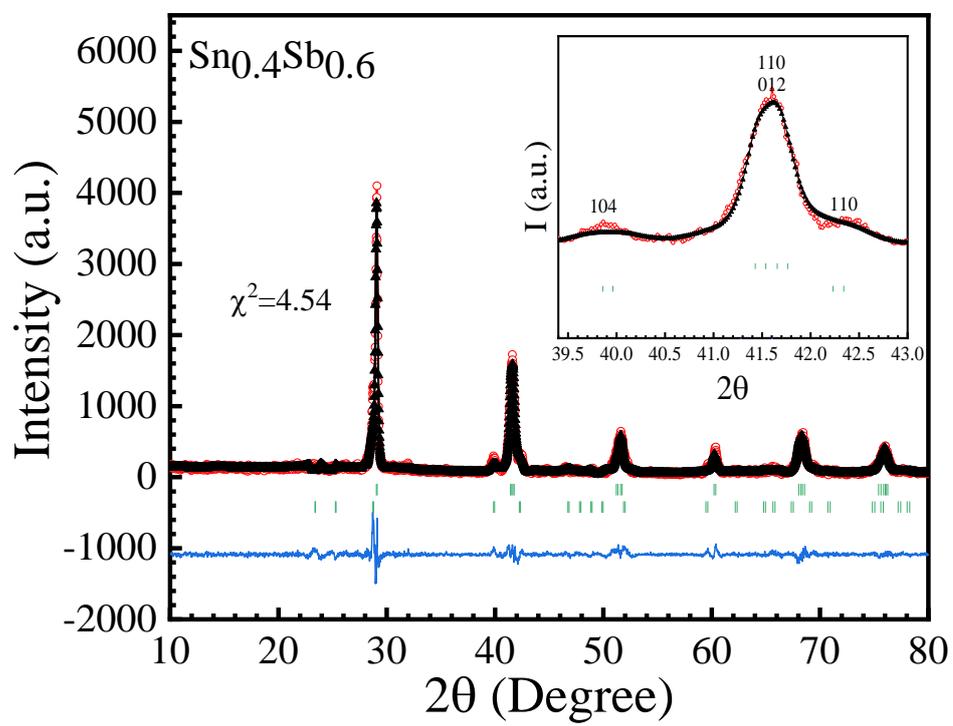

Fig. 2(c)

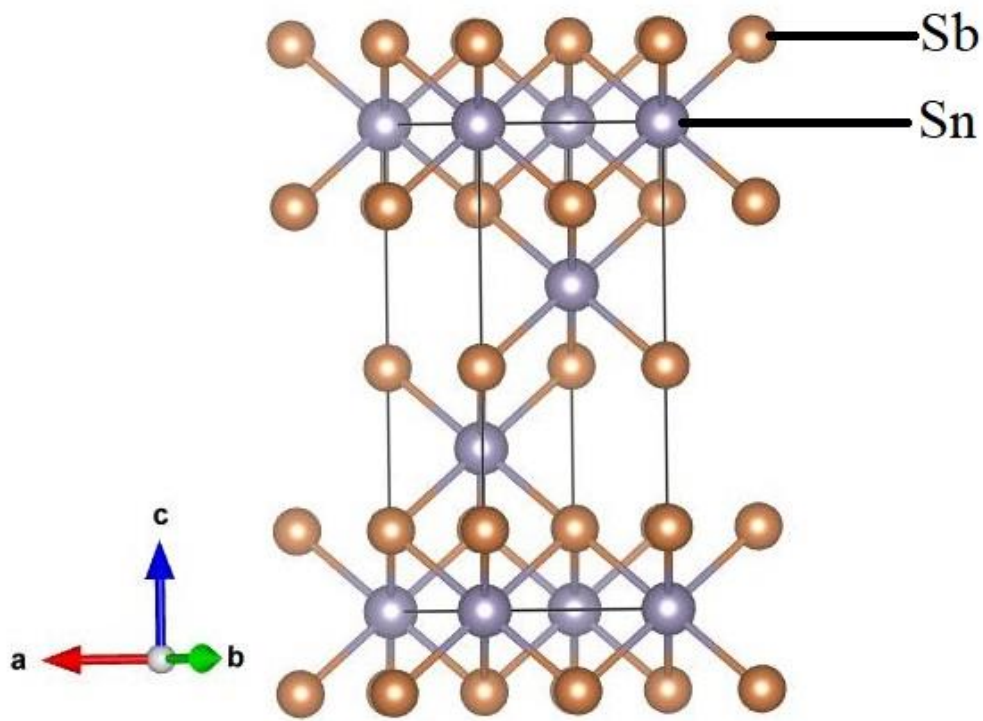

Fig. 3(a)

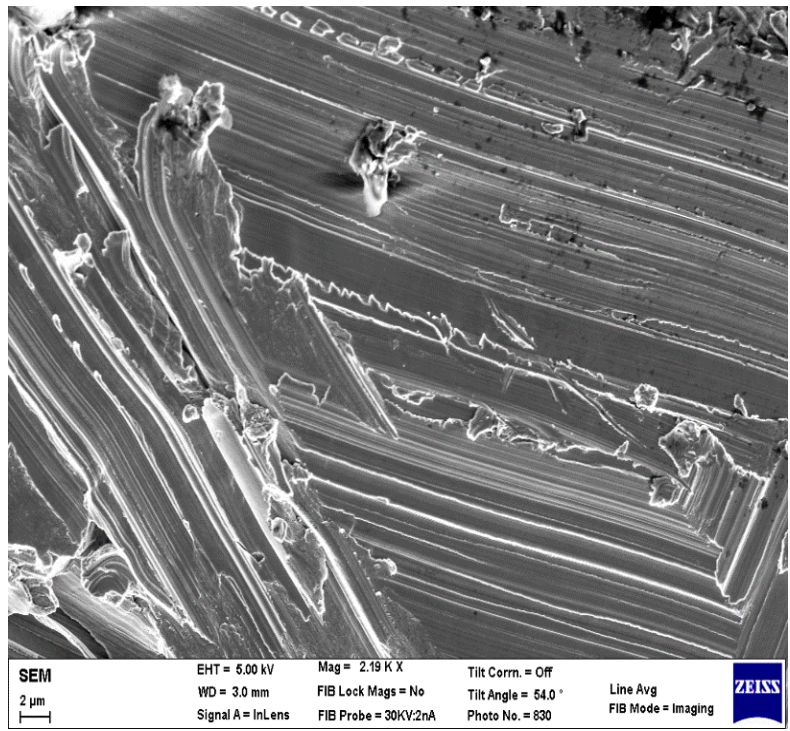

Fig. 3(b)

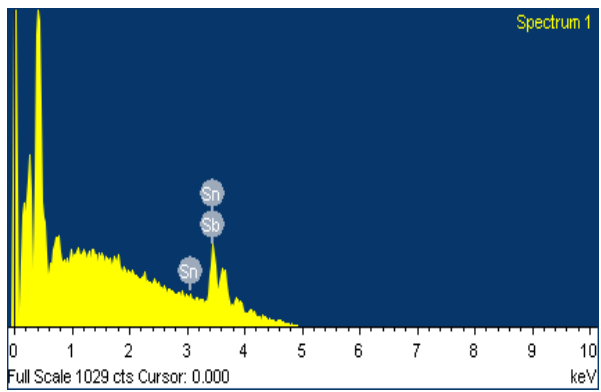

| Element | Weight % | Atomic % |
|---------|----------|----------|
| Sn L    | 50.16    | 50.80    |
| Sb L    | 49.84    | 49.20    |
| Totals  | 100.00   |          |

Fig. 4

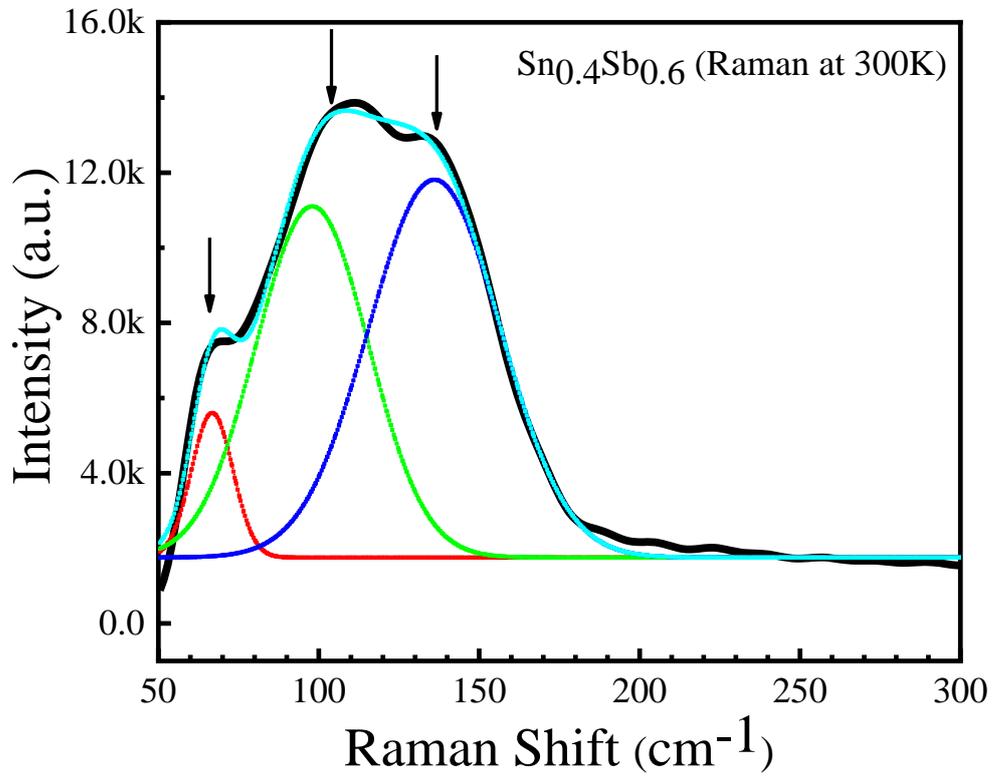

Fig.5

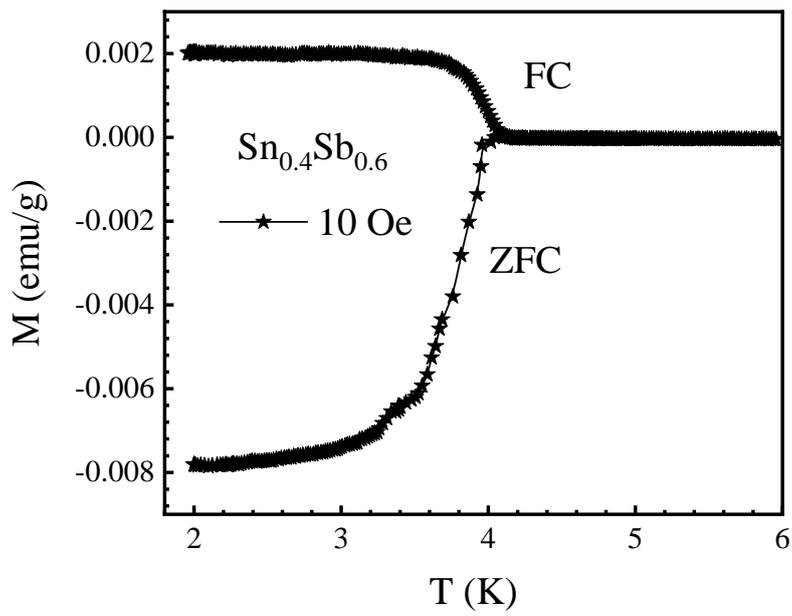

Fig. 6

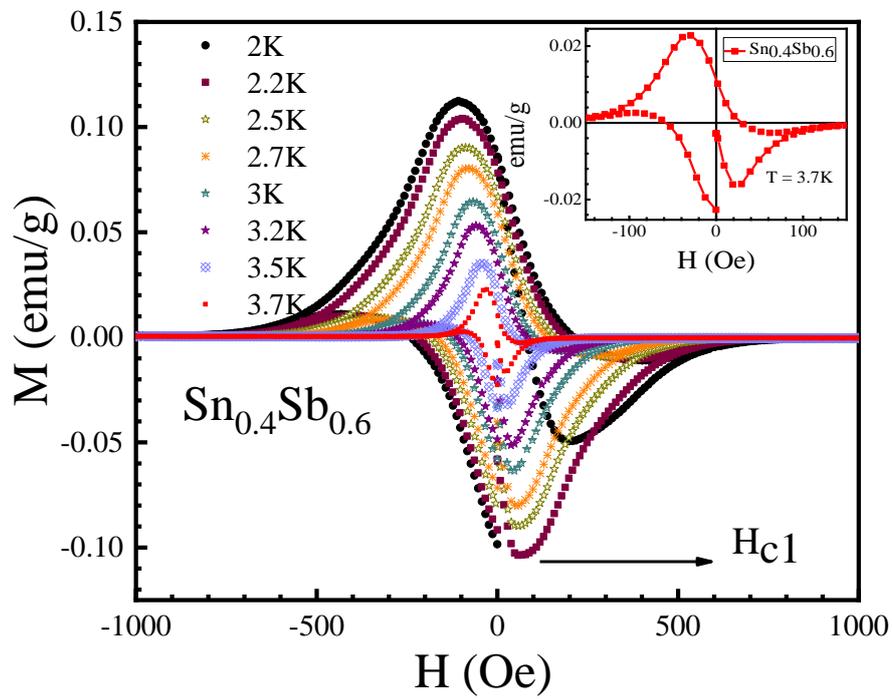

Fig. 7(a)

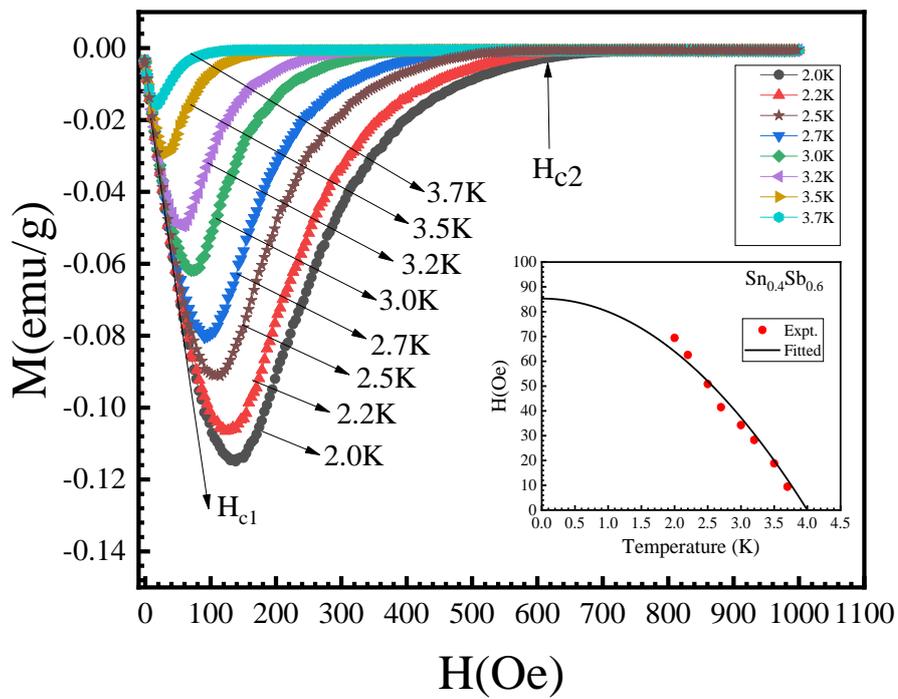

Fig. 7(b)

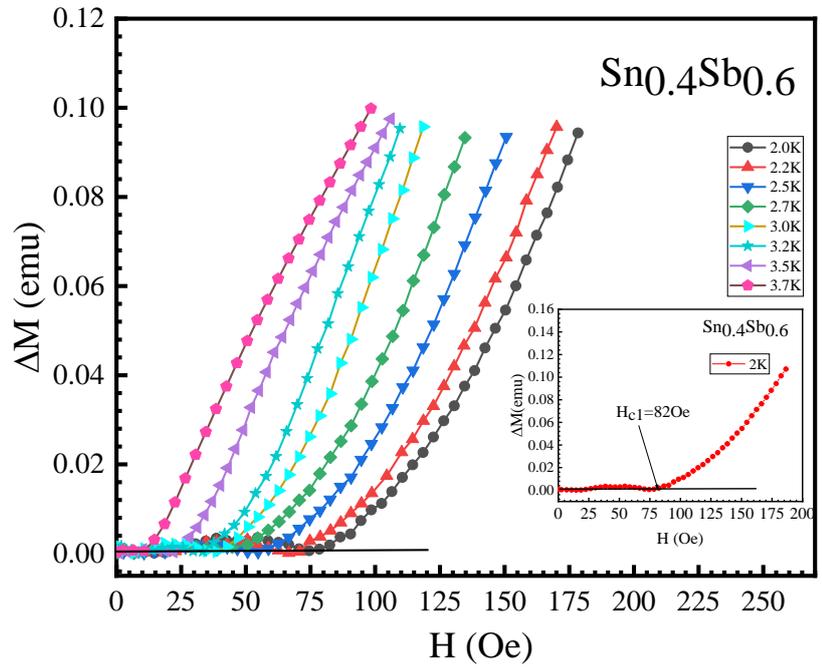

Fig. 8

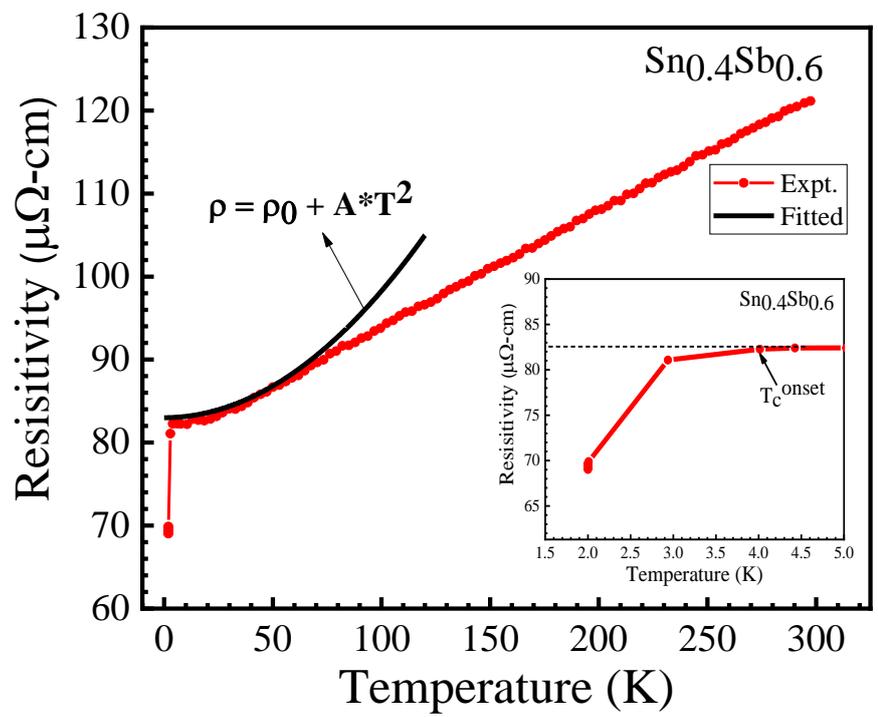